\documentclass[]{pasj01}
\draft

\begin{document} 
\Received{}
\Accepted{}

\title{Origin of cool cores, cold fronts and spiral structures in cool core clusters of galaxies}

\author{Hajime \textsc{Inoue}\altaffilmark{1}}%
\altaffiltext{1}{Institute of Space and Astronautical Science, Japan Aerospace Exploration Agency, 3-1-1 Yoshinodai, Chuo-ku, Sagamihara, Kanagawa 252-5210, Japan}
\email{inoue-ha@msc.biglobe.ne.jp}

\KeyWords{galaxies : clusters : general --- galaxies : clusters : intracluster medium --- X-rays : galaxies : clusters} 

\maketitle

\begin{abstract}
We consider a situation in which a brightest cluster galaxy (BCG) moves in ambient hot gas in the central region of a cool core cluster of galaxies, following the study by Inoue (2014, PASJ, 66, 60).
In the rest frame of the BCG, the hot gas is supposed to flow toward the BCG in parallel from a sufficiently large distance.
Then, it is expected that only the gas flowing with the impact parameter less than a critical value is trapped by the gravitation field of the BCG because of the efficient radiative cooling, getting a cooling flow, and that the remaining outer gas can get over the potential well.
In such a circumstance, we can draw the following picture:
A boundary layer between the out-flowing gas and the trapped gas arises around the stagnation point at the back side of the BCG.
Since the boundary temperature is so low as to be X-ray dim, the boundary could be observed as the cold front in X-rays.
The trapped gas once stagnates on the inner side of the boundary and starts in-falling toward the BCG.
Since the wandering motion of the BCG is likely to have a rotational component, the Coriolis force induces a rotational motion in the in-falling flow from the stagnation place to the BCG, forming a spiral structure around the BCG.
The spiraling flow turns the BCG on the up stream side of the main flow from the far outside, and arises another boundary layer having contact discontinuity with the main hot gas flow.
These pictures well reproduce the observed features such as cool cores, cold fronts, and spiral structures.
It can also be resolved how the cooling flow is suppressed from what the cooling flow hypothesis predicts, without any heating mechanism.

\end{abstract}

\section{Introduction}
Clusters of galaxies are divided into two families based on their core radii (Jones \& Forman 1984).
Clusters in one family have bright and cool cores with small radii and their X-ray emission is centered on a central dominant galaxy, which is called the brightest cluster galaxy (BCG) according to the related literature.
The clusters in this family are called cool core clusters.

Simple estimations from X-ray observations of the cool core clusters indicate that the central density of hot plasma in the clusters is so high for the hot plasma to radiatively cool down within the expected age of the cluster.  Then, the cooling gas is considered to flow inward to maintain a hydrostatic equilibrium with overlying gas; that is a cooling flow hypothesis (see Fabian 1994 for the review).

Careful and extensive studies of the central X-ray emission of the cool core clusters observed with ASCA (the Advanced Satellite for Cosmology and Astrophysics), however, showed that the results did not simply favor the cooling flow hypotheses: 
coexistence of hot (uncooled) and cool components even in the cluster core region, 
the smaller volume fraction of the cool component in the core than expected 
from the cooling flow hypothesis, 
differences in the chemical composition between the core region and the outer region, and so on (see Makishima et al. 2001 and references therein).

High resolution X-ray spectroscopic observations with XMM (X-ray Multi-mirror Mission) - Newton
further found that temperature of the hot plasma decreased toward 
the center but the cooling stopped at a certain temperature higher than 
what the cooling flow model predicts (Tamura et al. 2001; Kaastra et al. 2001; Peterson et al. 2001). 

These observations suggest presence of some heating mechanism at the center of the  clusters, and 
the AGN (active galactic nuclei) feedback is the currently favored one (see e.g. McNamara and Nulsen 2012 for the review).
The recent X-ray observation of the Perseus Cluster with the finest spectral resolution ever achieved with Hitomi, however, revealed that the velocity dispersion of the hot gas in the central region within $\sim$100 kpc was as low as $\sim$100 km s$^{-1}$ (Hitomi Collaboration 2016; 2018), implying that the low turbulent motion contradicted with the simple expectation from the AGN feedback scenario.

In parallel to the above studies, fine spatial structures have been resolved in the core regions with the finest angular resolution ever achieved of Chandra, such as sharp contact discontinuities (cold fronts) between regions with different temperatures (see Markevitch \& Vikhlinin 2007 for the review); spiral excess structures over the average image (Clarke et al. 2004; Blanton et al. 2011; Ghizzardi et al. 2014; Ichinohe et al. 2015; Ueda et al. 2017).

Markevitch, Vikhlinin and Mazzotta (2001) discussed that observed evidences of the cold fronts could be explained by a picture in which low entropy gas in the core was sloshing in the central potential well of the host cluster.
Then, Ascasibar and Markevitch (2006) performed detailed simulations, and showed that such structures as cold fronts and spiral structures could be reproduced by gas sloshing of the cluster's own cool, dense central gas as the result of a disturbance of the central potential by past subcluster in-fall.
Further simulations including radiative cooling done by ZuHone, Markevitch and Johnson (2010) revealed that sloshing could prevent a cooling catastrophe for intervals of time $\sim$1 - 3 Gyr, although eventually a cooling flow developed.
Thus, successive encounters with subclusters every 1 -3 Gyr are required for this process to significantly contribute to the cooling flow suppression.

Recently, Gu et al. (2020) proposed another mechanism for the cooling flow suppression in which some fraction of the dynamical energies of member galaxies moving in a cluster was transferred to the hot plasma through (magneto)hydrodynamic and gravitational interactions between them.
Although the estimated energy transfer rate to the hot plasma could be enough to prevent the cooling catastrophe, relations of the scenario with the cold fronts and spiral structures are unclear.

Interactions of the member galaxies with the central BCG was studied by Inoue (2014) and the results indicated that the BCG wandered with a velocity $\sim$ 100 km s$^{-1}$ around the cluster center.
Supposing a flow of the hot gas across the BCG with the relative velocity of  $\sim$ 100 km s$^{-1}$ in the rest frame of the BCG, Inoue (2014) investigated change of the specific energy of the hot gas when it passed by the BCG and found that a limited mass of the hot gas within a critical impact parameter was trapped by the gravitational potential of the BCG, forming the transverse cooling flow.

In this paper, we discuss how this transverse cooling flow scenario can well reproduce the observed structures and properties of the core region.

\section{Structures in the transverse cooling flow}
We consider a spherical region with radius, $R_{\rm c}$, at the center of a cool core cluster of galaxies, within which only one member galaxy exists on average.
Following the study by Inoue (2014), we assume that $R_{\rm c} \simeq$ 80 kpc and that the gravitational field of the BCG is dominant to that of the diffuse dark matter in this region.
Then, we expect a situation in which a BCG moves relatively to the ambient hot gas with velocity $\sim$ 100 km s$^{-1}$, as discussed by Inoue (2014).
Hereafter, we trace the flow of the hot gas in the gravitational field of the BCG in the rest frame of the BCG.

The wandering of the BCG is the result of successive encounters with the closest member galaxy which comes from a random direction in turn on the time scale of several $\times 10^{7}$ yrs.
The dynamical interaction with the closest member galaxy induces a rotational motion of the BCG around the barycenter.  
The specific angular momentum carried by the closest member galaxy should be $b v_{\infty}$ in terms of the impact parameter, $b$, and the incident velocity, $v_{\infty}$,  and the average rotational velocity, $\Omega$, over the passage of the closest member galaxy through the central region could roughly be estimated as
\begin{equation}
\Omega \sim \frac{b v_{\infty}}{R_{\rm c}^{2}},
\label{eqn:Omega}
\end{equation} 
considering that the staying time of the closest member galaxy in the relevant region would be the longest when the distance to the BCG is $\sim R_{\rm c}$.
The BCG should move with the same angular velocity around the barycenter, equation (\ref{eqn:Omega}) gives us the approximate angular velocity of the BCG.
Hence, rotational motions are likely to be induced in the hot gas flow by the Coriolis force in the BCG rest frame.

\begin{figure}
 \begin{center}
  \includegraphics[width=14cm]{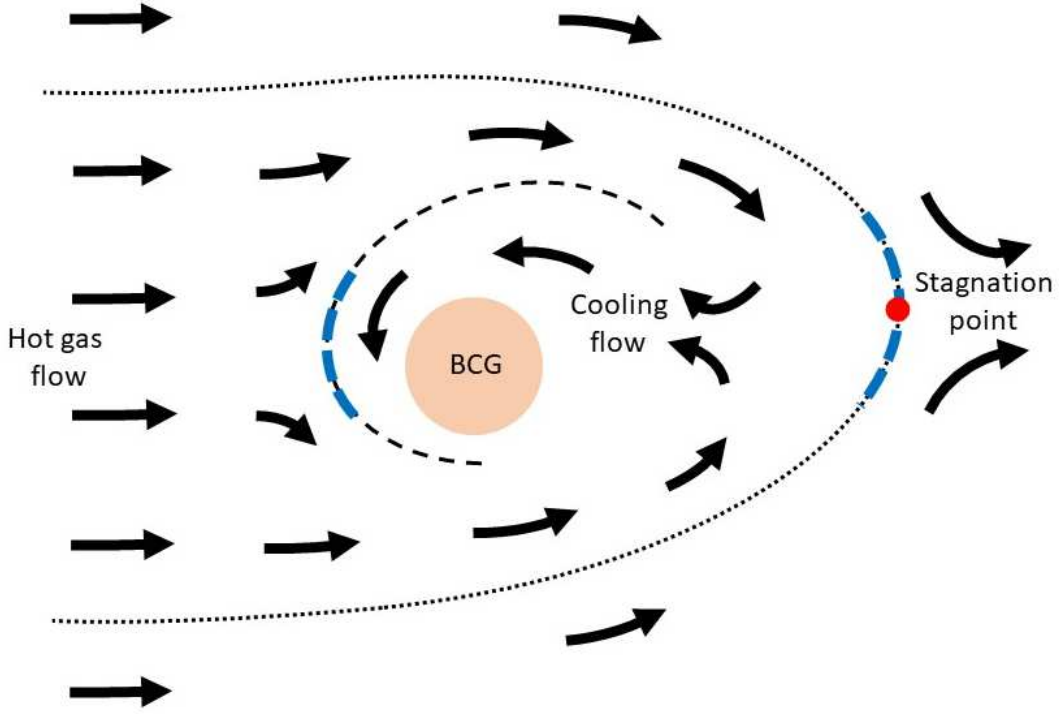}
 \end{center}
\caption{Configurations of hot gas flows around the BCG.  The outer dotted line indicates the boundary between the out-flowing gas and the trapped gas by the BCG, and the inner thin dashed line is the boundary with contact discontinuity between the hot gas flow towards the stagnation region and the cooling flow toward the BCG.  The thick dashed lines are the candidate places for the cold fronts.}
\label{fig:Configurations}
\end{figure}

The flow configurations considered here are schematically drawn in figure \ref{fig:Configurations}.

\subsection{Outer cold front}\label{OuterColdFront}
We suppose that the hot gas flows towards the BCG in parallel from an outer boundary far from the BCG.
In appendix, the flow properties are obtained for a simplified case without radiative cooling nor effect of the Coriolis force.
Then, the effect of the radiative cooling is estimated for the simplified flow. 
The results of appendix show that when the hot gas density is higher than a critical one, the radiative cooling is so efficient for the flow with an impact parameter less than a critical value that the energy becomes insufficient to get over the potential well of the BCG.

\begin{figure}
 \begin{center}
  \includegraphics[width=14cm]{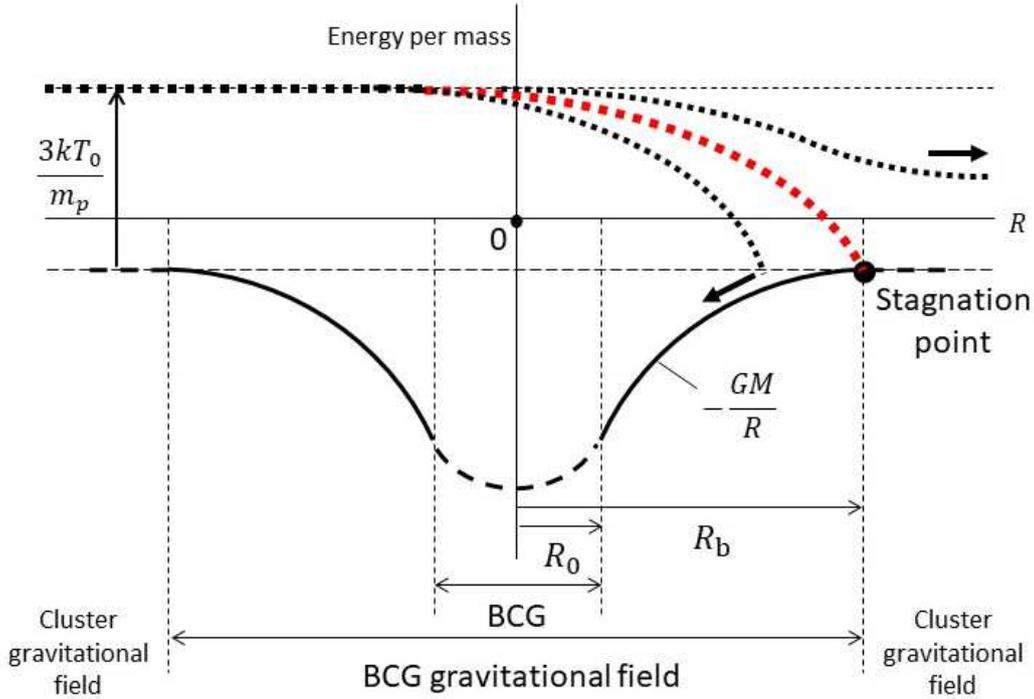}
 \end{center}
\caption{Schematic diagram for specific energy variations of the gas flowing across the BCG gravitational potential, as a function of the distance, $R$, of a gas position from the BCG center.  The thick dotted line is the boundary between the flow trapped by and the flow escaping from the potential.}
\label{fig:SpecificEnergy}
\end{figure}

If this really happens, a boundary is expected to appear along the flow with the critical impact parameter, outside of which the gas is possible to flow out but inside of which the gas is trapped in the potential well.
Figure \ref{fig:SpecificEnergy} shows the schematic trajectory of the specific total energy of the gas moving along the boundary across the BCG.
This boundary is considered to have a stagnation point between the outflowing gas and the inflowing gas, which should locate at the edge of the region dominated by the gravitational potential of the BCG, as expressed with the filled circle  in figure \ref{fig:SpecificEnergy}.
The outer region than the BCG-dominant region is governed by the gravitational field of the cluster dark matter or the nearest member galaxy.
As seen from this figure, the specific total energy should be just the potential energy of the BCG at the stagnation point.
The specific kinetic energy of the bulk motion at the stagnation point should be zero by definition.  
The specific thermal energy is also required to be zero there, otherwise remaining thermal energy should induce outward bulk motion.
Hence, it is expected that the gas temperature is so low for the boundary near the stagnation point to be dim in the X-ray band and that the temperature has an apparent jump across the boundary in the X-ray observation, although the physical quantities smoothly varies across the boundary.

The gas trapped by the BCG inside the outer boundary surface is generally considered to change its flow direction from outward to inward at some distance, $R$.
The temperature of the gas at the flow reversing position, $T_{\rm r}$, is roughly given as
\begin{equation}
T_{\rm r} \simeq T_{\rm b} \left( \frac{R_{\rm b}}{R} - 1 \right),
\label{eqn:Tr}
\end{equation}
where
\begin{eqnarray}
T_{\rm b} &=& \frac{m_{\rm p}GM}{3 k R_{\rm b}} \nonumber \\
&=& 2.2 \times 10^{7} \left(\frac{M}{10^{13} M_{\odot}}\right) \left(\frac{R_{\rm b}}{80 \mbox{ kpc}}\right)^{-1} \rm{K},
\label{eqn:Ts}
\end{eqnarray}
and $R_{\rm b}$ is the distance of the edge of the BCG gravitational field.
The gas is assumed to be pure hydrogen, and $m_{\rm p}$, $G$, $M$, and $k$ are the proton mass, the gravitational constant, the mass of BCG and the Boltzmann constant respectively.
According to the estimation of $R_{\rm b} \sim$ 70 kpc by Inoue (2014), we assume $R_{\rm b} \sim R_{\rm c} \sim$ 80 kpc here.
Since $R$ is likely to distribute from near the outer boundary of the BCG, $R_{0}$ (assumed to be $\sim$20 kpc here), to $R_{\rm b}$, the observationally averaged temperature of the flow-reversing region could be $\sim 1 keV$.

The temperature of the outflowing gas outside the outer boundary surface should observationally biased to the intrinsic gas temperature, $T_{0}$ ($\sim 10^{8}$ K), considering integration of the flux over the line of sight, and is expected to be observationally higher than that observed from the inner side of the outer boundary.  

Since the relatively cold gas sits on the inner side viewed from the BCG, we can call this boundary the cold front.  To distinguish it from the other one discussed in subsection \ref{InnerCF}, this cold front is named the outer cold front.

\subsection{Cooling flow with spiral structure}
The cool gas in the flow-reversing region should start in-falling toward the BCG, getting the transverse cooling flow. 
Since the Coriolis force in the BCG frame induces rotational motions, the cooling flow is expected to form a spiral structure around the BCG.

The angular velocity of the BCG rest frame, $\Omega$ is roughly estimated in equation (\ref{eqn:Omega}).
Since we can approximately consider that the starting point of the cooling flow rotate with $\Omega$ around the BCG, the specific angular momentum of the in-falling gas, $\ell$, is roughly given as
\begin{equation}
\ell \sim R_{\rm r}^{2} \Omega,
\label{eqn:ell}
\end{equation}
where $R_{\rm r}$ is the average distance of the reversing position of the gas within in the flow-reversing region.
Then, the in-flowing gas tends to rotate around the Keplerian circular orbit determined by $\ell$ and its radius, $R_{\rm k}$, is calculated with the help of equations (\ref{eqn:ell}) and (\ref{eqn:Omega}) as
\begin{eqnarray}
R_{\rm k} &=& \frac{\ell^{2}}{GM} \nonumber \\ 
&\sim& \frac{R_{\rm r}^{4}b^{2}v_{\infty}^{2}}{R_{\rm c}^{4} GM}.
\label{eqn:R_k}
\end{eqnarray}
Considering $GM/v_{\infty}^{2} \simeq$ 40 kpc for $v_{\infty} \simeq 10^{8}$ cm s$^{-1}$ (typical velocity of the member galaxy) and $M \simeq 10^{13} M_{\odot}$, $R_{\rm c} \sim$ 80 kpc and that the average impact parameter of the innermost member galaxy, $b$, should be slightly smaller than $R_{\rm c}$, we could roughly have a relation as
\begin{equation}
\frac{b^{2}v_{\infty}^{2}}{R_{\rm c}GM} \sim 1.
\label{eqn:Simplification}
\end{equation}
Then, we have an approximate equation as
\begin{equation}
R_{\rm k} \sim \left(\frac{R_{\rm r}}{R_{\rm c}}\right)^{3} R_{\rm r}.
\label{eqn:R_k-R_c}
\end{equation}
If $R_{\rm r}$ is several 10\% of $R_{\rm b} \sim R_{\rm c}$, $R_{\rm k}$ is roughly estimated to be a few 10\% of $R_{\rm r}$ and a significant rotational motion is expected to appear in the cooling flow.

The schematic diagram of the spiral cooling flow as viewed from top is exhibited in figure \ref{fig:Configurations}.

\subsection{Inner cold front}\label{InnerCF}
As discussed above, the cooling gas in-flows from the flow-reversing region and tends to turn around the BCG on the up stream side of the main gas flow.
Then, the in-flowing gas should interact with the hot gas in the main flow, and could lose the angular momentum to eventually be absorbed in the atmosphere of the BCG.

As shown in the schematic picture in figure \ref{fig:Configurations}, an elongated surface with contact discontinuity is expected to appear around the turning cooling flow, where the cooling flow and the main flow have opposite flow-directions to each other.
At the most up-stream side of the boundary surface viewed from the main flow, a stagnation point could emerge in the main flow and another cold front is expected to appear there.
We call this boundary the inner cold front because of the closer position to the BCG than the outer cold front.
The distance of the inner cold front is considered to be $\sim R_{\rm k}$ and be  roughly a few 10\% of the typical size of the flow-reversing region as discussed in the previous subsection.

\section{Comparison with observations}
Considering the situation in which the hot gas flows toward the BCG in parallel from the outer boundary far from the BCG, the study in appendix indicates that when the hot gas density is higher than a critical one, the radiative cooling is so efficient for the flow with an impact parameter less than a critical value that the energy becomes insufficient to get over the potential well of the BCG.
Then, the gas trapped in the gravitational potential of the BCG is discussed to form the outer cold front, the flow-reversing region, the spiral structure and the inner cold front, in the previous section.
The overall configurations of such structures, as schematically drawn in figure \ref{fig:Configurations}, look consistent with observations of cool core regions of several cool core clusters of galaxies (e.g. Clark et al. 2004; Sanders et al. 2005; Ghizzardi et al. 2014; Ichinohe et al. 2015).

According to appendix, the critical density above which the energy-exhaustion of the flow occurs within the relevant region is $\sim$ 0.3 $n_{\rm p,\; 00}$ and is $\sim$ 0.1 cm$^{-3}$ for the $n_{\rm p,\; 00}$ value from equation (\ref{eqn:n00}) for $M = 10^{13} M_{\odot}$, $R_{0}$ = 20 kpc and $v_{\rm b} = 10^{7}$ cm s$^{-1}$.
The density discussed in appendix, $n_{\rm p,\; 0}$ is the hot gas density at the boundary with the BCG, while the density in the cool core region should be lower than $n_{0}$ by factor of 2 to 3.
Furthermore, the estimation of the critical density depends on the simplified case of the isothermal flow without energy loss.
In reality, the temperature should gradually decreases, inducing a contraction of the flow cross section, as the flow advances, and thus the effect of the radiative cooling should be more enhanced. 
From these considerations, we can  say that the number density of the order of 10$^{-2}$ cm$^{-3}$ observed from the cool core regions of the cool core clusters is consistent with the present scenario.

The presence of two boundaries, the inner and outer one, is predicted in the present scenario.
The outer one is the boundary around the stagnation point where parallel flow from the up-stream side separates into outward flow and inward flow.
Although no real contact discontinuity exists, jumps of the spectral parameters could be seen across the boundary in the X-ray band since the temperature around the stagnation point is too low for X-rays to be observed from the vicinity of the boundary.
The inner one is, on the other hand, the boundary between the cooling gas inflowing from the flow-reversing region behind the outer boundary and the main flow from the far up-stream side, and the contact discontinuity is expected to really exist. 
Observationally, more than one cold front are found in several clusters of galaxies (see e.g. Ghizzardi et al. 2010).
In particular, Ascasibar and Markevitch (2006) shows clear presence of two, inner and outer edges in some sources.
The inner edges seen in figure 1 in Ascasibar and Markevitch (2006) look clearer than the outer edges, which could be consistent with the above difference in the boundary property.
The difference in the degree of the temperature jump between the outer cold front and the inner front is also seen in figure 6 of Sanders, Fabian and Taylor (2005).

As discussed in the previous section, the distance of the outer boundary is considered to be several ten kpc, while that of the inner boundary is smaller by a factor of a few.
The distances of the cold fronts obtained from observations by Ghizzardi, Rossetti and Molendi (2010)  are consistent with this expectation (see figure 4 in their paper, excluding the results from merging clusters).

The flow directions along a part of the inner boundary are opposite to each other.  This could cause the Kelvin-Helmholtz instability there as suggested from the observation by Ghizzardi, De Grandi and Molendi (2014).

The gas trapped by the BCG starts spiraling in from the region inside the outer cold front to the region inside the inner cold front, finally merging with the BCG.
Since the efficient radiative cooling causes the cooling flow, the flow should be observed as the bright spiral structure in X-rays as several clusters exhibit (Clarke et al. 2004; Blanton et al. 2011; Ghizzardi et al. 2014; Ichinohe et al. 2015; Ueda et al. 2017).
The gas temperature should gradually decrease along the in-falling flow, and thus that inside the inner cold front is likely to be lower than that inside the outer cold front.
In fact, the temperatures on the cold sides of the the outer and inner cold fronts analyzed by Sanders, Fabian and Taylor (2005) agree to this inference. 

The cooling flow is considered to flow along the outer rim of the BCG, and thus mixing of the in-falling gas with the matter in the BCG can be expected to arise.
This is consistent with the observed metal abundance maps of the cool core regions too (see e.g. figure 9 in Ghizzardi et al. 2014).

As discussed in subsection \ref{OuterColdFront}, the gas around the stagnation point has a very low temperature and is likely to emit H$\alpha$ lines.
Indeed, H$\alpha$ filaments are observed from several cool core clusters of galaxies (McDonald et al. 2010; 2011; 2012).
McDonald et al. (2010) and McDonald, Veilleux and Mushotzky (2011) showed that the 0.5-2.0 keV X-ray and H$\alpha$ morphologies are correlated and that H$\alpha$ filaments extend all the way to the radius of the X-ray cool core but never beyond.
This agrees to the discussion that the inward gas flow emitting H$\alpha$ lines starts from the stagnation point (outer edge of the cool core region) and the gas emitting soft X-rays flows from the vicinity to the stagnation point along with the H$\alpha$ emitting gas.
McDonald, Veilleux and Rupke (2012) revealed that the widths of the emission lines in the H$\alpha$ filaments decrease with radius, from FWHM $\sim$600 km s$^{-1}$ in the central region to FWHM $\sim$100 km s$^{-1}$ in the most extend filaments and that the kinematic signatures of the lines are consistent with rotation.
This strongly supports the picture of the transverse cooling flow under the effect of the Coriolis force as discussed in the previous section.

\section{Discussion}
The present scenario predicts that when a BCG moves through ambient hot gas in the central region of a cool core cluster, a transverse cooling flow appears in association with such structures as discussed above.
We need no additional members, such as the repeatedly in-falling subclusters, to the intrinsic constituents of the cool core clusters: the BCG, the member galaxies, the hot gas and the diffuse dark matter. 
No other activities, such as the AGN feedback, is either needed than the mutual dynamical interactions between the constituents.

The mechanism for wandering of a BCG was studied by Inoue (2014):
The encounter between the BCG and the innermost member galaxy induces a movement of the central object, trying to establish the equipartition of their kinetic energies.
If the BCG starts moving, however, 
the ambient dark matter  brake the moving through the dynamical friction.
Then, a balance between the energy gain rate of the BCG 
from the innermost galaxy passing by and the energy loss rate 
through the dynamical friction to the diffuse dark matter determines the wandering velocity of the BCG.

The energy is also transfered from the innermost member galaxy to the hot gas in the central region of the cluster, as discussed by Gu et al. (2020).
The energy transfer rate through the dynamical friction is proportional to $M_{\rm mg}^{2}/v_{\rm mg}$ according to Ostriker (1999), and that through the ram pressure could be $\propto \rho_{\rm mg} R_{\rm mg}^{2} v_{\rm mg}^{3}$, where  $M_{\rm mg}$, $\rho_{\rm mg}$, $R_{\rm mg}$ and $v_{\rm mg}$ are the average mass, density, and size of the member galaxies and the relative velocity of the member galaxy to the hot gas, respectively.
On the other hand, if the BCG with mass, $M_{\rm bcg}$, density, $\rho_{\rm bcg}$ and size, $R_{\rm bcg}$, moves relatively to the hot gas with velocity, $v_{\rm bcg}$, the energy transfer rates from the BCG to the hot gas through the dynamical friction and through the ram pressure are proportional to $M_{\rm bcg}^{2}v_{\rm bcg}^{2}/c_{\rm s}^{3}$ (Ostriker 1999) and to $\rho_{\rm bcg}R_{\rm bcg}^{2}v_{\rm bcg}^{3}$,  respectively.  Here, $c_{\rm s}$ is the sound velocity of the hot gas.
Assuming $M_{\rm mg} \sim 10^{12} M_{\odot}$, $M_{\rm bcg} \sim 10^{13} M_{\odot}$, $\rho_{\rm mg} \sim \rho_{\rm bcg}$, $R \propto M^{1/3}$,  $v_{\rm mg} \sim c_{\rm s} \sim 10^{8}$ cm s$^{-1}$ and $v_{\rm bcg} \sim 10^{7}$ cm s$^{-1}$, the energy transfer rates through the dynamical friction from the member galaxy and from the BCG are comparable to each other, but that through the ram pressure from the member galaxy is much larger than that from the BCG.  
Thus, the dynamical situation of the hot gas is expected to be mainly governed by the interactions with the innermost member galaxy and to be independent of the motion of the BCG on average.

Under such circumstances, interaction between the wandering BCG and the ambient hot gas and appearance of the transverse cooling flow are discussed by Inoue (2014).  

Since only the matter flowing with the impact parameter less than a critical value is involved in the transverse cooling flow, the matter flow rate of the cooling flow is largely suppressed from that in the case of the radial cooling flow.
This is simply due to the following reason.

The density distribution of the hot gas is determined by the hydrostatic equation in the gravitational field of the BCG and the density decreases as the distance from the BCG increases.  Even if the hot gas moves relatively to the BCG as considered here, the velocity is sufficiently subsonic for the hydrostatic structure to be assured.
Then, in the case of the canonical radial cooling flow, the hot gas having a density larger than a critical one with which the cooling time equals to the age of the cluster, $\sim$10 Gyr, should be involved in the cooling flow.
However, in the case of the transverse cooling flow, the critical density is determined in terms of $n_{\rm p,\; 00}$ defined in equation (\ref{eqn:n00}), with which the radiative cooling time of the hot gas with temperature $m_{\rm p}GM/(kR_{0})$ equals to the time for the hot gas with velocity $v_{\rm b}$ to go across the distance $R_{0}$.
Since $R_{0}/v_{\rm b} \sim $ 0.2 Gyr for $R_{0}$ = 20 kpc and $v_{\rm b}$ = 100 km s$^{-1}$, 
the region of hot gas involved in the cooling flow is expected to be largely suppressed in the transverse case from that in the radial case.

As described above, the present scenario can well explain several observed properties of the cool core region.
They can be understood as results of the ordinary and passive interactions between the constituents in the cool core clusters of galaxies.
It is consistent with the low turbulent situation of the cool core region indicated from the Hitomi observation, and the common presence of the cold fronts to the cool core regions inferred from the statistical study by Ghizzardi, Rossetti and Molendi  (2010).
Detailed simulations on this scenario are desired.



\appendix 
\section*{Effect of radiative cooling on a parallel flow passing by the BCG}
We introduce a cylindrical coordinate ($r$, $\phi$, $z$) and set the origin at the center of the BCG.
We consider a steady flow of the hot gas from the minus $z$ side to the plus $z$ side which is axial-symmetric around the $z$ axis with $r = 0$ and neglect the effects of the radiative cooling and the Coriolis force for simplicity here.
The hot gas is assumed to inflow from the inflowing boundary surface at $z = -z_{\rm b}$ with a constant velocity, $v_{\rm b}$, which is sufficiently smaller than the sound velocity of the hot gas.

In the sufficiently subsonic case, the proton number density of the hot gas, $n_{\rm p}$, is approximately calculated by the hydrostatic equation as
\begin{equation}
\frac{dP}{dR} = -n_{\rm p}\; m_{\rm p} \frac{GM}{R^{2}},
\label{eqn:dPdR}
\end{equation}
where $P$, $m_{\rm p}$, $G$, and $M$ are the pressure of the hot gas, the proton mass, the gravitational constant and the mass of the BCG respectively, and we assume that the hot gas is pure hydrogen gas.
$R$ is the distance of the hot gas position from the gravity center and is expressed as
\begin{equation}
R = \sqrt{r^{2} + z^{2}}.
\label{eqn:R}
\end{equation}
On the isothermal approximation with temperature, $T_{0}$, equation (\ref{eqn:dPdR}) is easily solved and the distribution of $n_{\rm p}$ is obtained as
\begin{equation}
n_{\rm p} = n_{\rm p,\; 0} \exp \left[ -\frac{1}{\tau} \left( 1 - \frac{R_{0}}{R} \right) \right],
\label{eqn:n}
\end{equation}
where $\tau$ is defined as
\begin{eqnarray}
\tau &=& \frac{2 k T_{0} R_{0}}{m_{\rm p} G M} \nonumber \\
&=& 0.77 \left(\frac{T_{0}}{10^{8}K}\right) \left(\frac{R_{0}}{20\; \rm{kpc}}\right) \left(\frac{M}{10^{13}M_{\odot}}\right)^{-1}.
\label{eqn:tau}
\end{eqnarray}
$R_{0}$ is the outer boundary of the BCG, $n_{\rm p,\; 0}$ is the proton number density at the boundary and $k$ is the Boltzmann constant.
Hereafter, we assume $\tau \simeq 1$.

If we neglect the energy loss of the flow, we can have the following Bernoulli's equation along a stream line as
\begin{equation}
\frac{v^{2}}{2} + \frac{2kT_{0}}{m_{\rm p}} \ln n_{\rm p} - \frac{GM}{R} = \frac{v_{\rm b}^{2}}{2} + \frac{2kT_{0}}{m_{\rm p}} \ln n_{\rm p,\; b} - \frac{GM}{R_{\rm b}},
\label{eqn:Bernoulli}
\end{equation}
where $v$ is the flow velocity and the parameters with the subscript $b$ in the right hand side are the respective values at the inflowing boundary.
Since we see
\begin{equation}
\frac{2kT_{0}}{m_{\rm p}} \ln n_{\rm p} - \frac{2kT_{0}}{m_{\rm p}} \ln n_{\rm p,\; b} = \frac{GM}{R} - \frac{GM}{R_{\rm b}},
\label{eqn:ln_n}
\end{equation}
with the help of equation (\ref{eqn:n}), 
we get from equation (\ref{eqn:Bernoulli})
\begin{equation}
v = v_{\rm b},
\label{eqn:v=v_b}
\end{equation}
indicating that the flow velocity is constant everywhere along the stream line.

\begin{figure}
 \begin{center}
  \includegraphics[width=12cm]{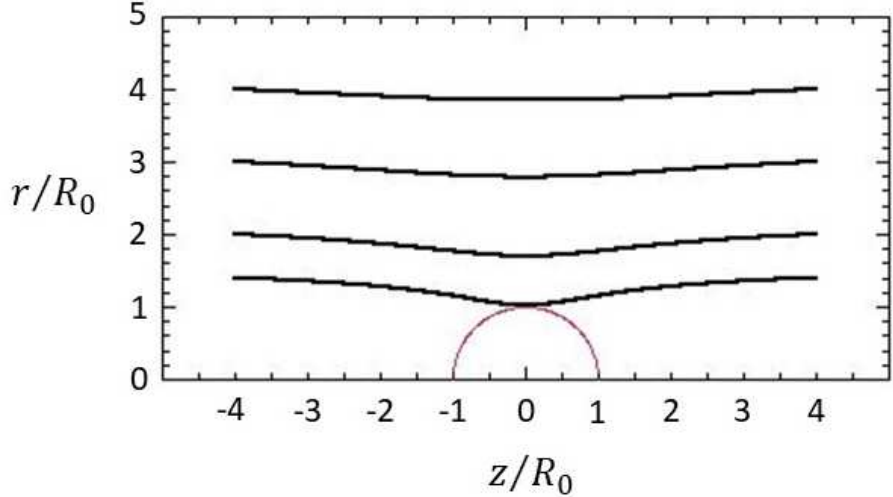}
 \end{center}
\caption{Four examples of stream lines obtained with equation (\ref{eqn:drdz}).  The central circle indicates the boundary of the BCG.  The coordinates are normalized by its radius.}
\label{fig:StreamLines}
\end{figure}

Considering stream lines in a cylinder-like shell with radius $r \sim r + \Delta r$,
the continuity equation is given as
\begin{equation}
nv\Delta S = \rm{constant},
\label{eqn:ContEq}
\end{equation}
where 
\begin{equation}
\Delta S = 2 \pi r \Delta r.
\label{eqn:DeltaS}
\end{equation}
Since $v$ is constant, we can approximately get from equation (\ref{eqn:ContEq})
\begin{equation}
\frac{1}{n} \frac{\partial n}{\partial z} + \frac{1}{\Delta S}\frac{d\Delta S}{dz} = 0.
\label{eqn:DiffContEq}
\end{equation}
Substituting equations (\ref{eqn:n}) and (\ref{eqn:DeltaS}) into equation (\ref{eqn:DiffContEq}) and considering $\Delta r \propto r$, we obtain an equation to determine the locus of the stream line as
\begin{equation}
\frac{dr}{dz} = \frac{1}{2\tau} \frac{r z R_{0}}{(r^{2} + z^{2})^{3/2}}.
\label{eqn:drdz}
\end{equation}
Some examples of the obtained stream lines from this equation are shown in figure \ref{fig:StreamLines}.
The stream lines are slightly constricted in the middle around the BCG but have a almost straight cylindrical shape.

\begin{figure}
 \begin{center}
  \includegraphics[width=10cm]{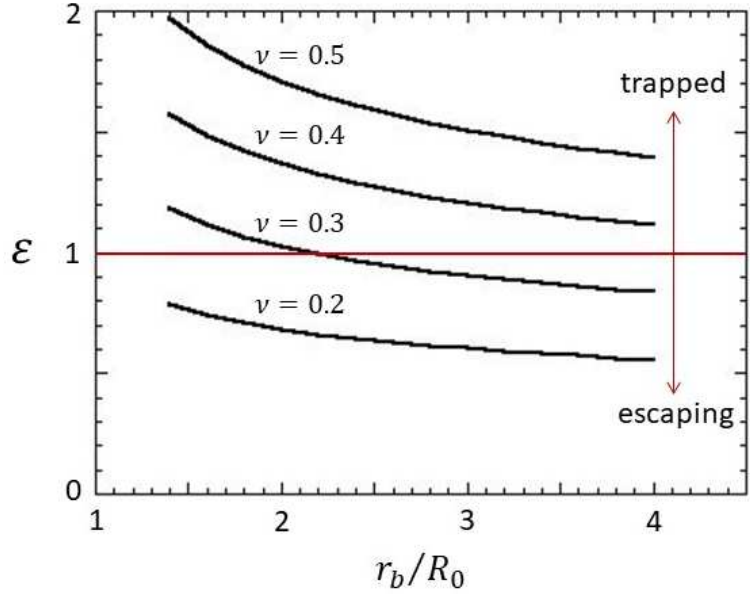}
 \end{center}
\caption{Ratios of the total specific energy loss integrated from $z = - 4\; R_{0}$ to $z = 4\; R_{0}$ to the initial net specific energy at the inflowing boundary at $z = -4\; R_{0}$ in the four cases of $\nu$.  The hot gas in the region above the $\varepsilon = 1$ line can be considered to be trapped by the BCG.}
\label{fig:RadiativeLoss}
\end{figure}

Now, we estimate the effect of the radiative energy loss on the flow.
Let $E$ be the net specific energy of the flowing hot gas, then 
the gas should lose its energy along the stream with approximate rate per unit length as
\begin{equation}
\frac{dE}{dz} = -\frac{n_{\rm p} \Lambda}{m_{\rm p} v},
\label{eqn:dEdz}
\end{equation}
where $\Lambda$ is the cooling function of thermal radiation (see e.g. Sutherland \& Dopita  1993).
With this equation, we can calculate the total specific energy loss of the gas, $\Delta E$, if it moves over the stream line from $z=-z_{\rm b}$ to $z=z_{\rm b}$, as
\begin{equation}
\Delta E = \frac{\Lambda}{m_{\rm p}v_{\rm b}} \int_{-z_{\rm b}}^{z_{\rm b}} n_{\rm p} dz,
\label{eqn:DeltaE}
\end{equation}
where the integration is done along the stream line.
Here, the net specific energy at the inflowing boundary, $E_{\rm b}$, is 
\begin{eqnarray}
E_{\rm b} &\simeq& \frac{3kT_{0}}{m_{\rm p}} - \frac{GM}{R_{\rm b}} \nonumber \\ &=& \left( \frac{3}{2} \tau - \frac{R_{0}}{R_{\rm b}}\right) \frac{GM}{R_{0}},
\label{eqn:E_b}
\end{eqnarray}
neglecting the kinetic energy because of the largely subsonic situation, and it is positive for $\tau \simeq 1$.
If $\Delta E$ exceeds $E_{\rm b}$, $E$ turns to negative and the hot gas becomes unable to escape from the potential of the BCG.
Then, the ratio, 
\begin{equation}
\varepsilon = \frac{\Delta E}{E_{\rm b}}
\label{eqn:Epsilon_Def}
\end{equation}
is calculated as
a function of $r_{\rm b}$, the impact parameter at the inflowing boundary, for several cases of a parameter, $\nu$, defined as
\begin{equation}
\nu = \frac{n_{\rm p,\; 0}}{n_{\rm p,\; 00}},
\label{eqn:nu}
\end{equation}
where
\begin{eqnarray}
n_{\rm p,\; 00} &=& \frac{m_{\rm p}GM v_{\rm b}}{R_{0}^{2} \Lambda} \nonumber \\
&=& 2.9 \times 10^{-1} \left(\frac{M}{10^{13}M_{\odot}}\right) \left( \frac{R_{0}}{20\; kpc}\right)^{-2} \left(\frac{v_{\rm b}}{10^{7} \mbox{cm s}^{-1}}\right) \left( \frac{\Lambda}{\Lambda_{8}}\right)^{-1} \mbox{ cm}^{-3}.
\label{eqn:n00}
\end{eqnarray}
$\Lambda_{8}$ is the cooling function for plasma with temperature of 10$^{8}$ K and the cosmic abundance, and is assumed to be  $2 \times 10^{-23}$ erg cm$^{3}$ s$^{-1}$ taken from Sutherland and Dopita (1993).
Figure \ref{fig:RadiativeLoss} shows the results in four cases of $\nu$ from 0.2 to 0.5.
As seen from this figure, the $\nu$ curve comes to cross the $\varepsilon=1$ line in the region of $r_{\rm b} > R_{0}$, when $\nu$ exceeds 0.3, and the $r_{\rm b}$ value at the cross point increases as $\nu$ increases.
This indicates that when the hot gas density is larger than a critical value, the effect of the radiative cooling becomes so large that the hot gas inflowing with impact parameter less than a critical one gets trapped by the gravitational potential of the BCG.

\begin{thebibliography}{}
\bibitem[]{}
Ascasibar, Y., \& Markevitch, M.\ 2006, \apj, 650, 102
\bibitem[]{}
Blanton, E.L., Randall, S.W., Clarke, T.E., Sarazin, C.L., McNamara, B.R., Douglass, E.M., \& McDonald, M.\ 2011, \apj, 737, 99
\bibitem[]{}
Clarke, T.E., Blanton, E.L., \& Sarazin, C.L.\ 2004, \apj, 616, 178
\bibitem[]{}
Fabian, A.C.\ 1994, \araa, 32, 277
\bibitem[]{}
Ghizzardi, S., De Grandi, S., \& Molendi, S.\ 2014, \aap, 570, A117
\bibitem[]{}
Ghizzardi, S., Rossetti, M., \& Molendi, S.\ 2010, \aap, 516, A32
\bibitem[]{}
Gu, L. et al.\ 2020, \aap, 638, A138
\bibitem[]{}
Hitomi Collaboration\ 2016, \nat, 535, 117
\bibitem[]{}
Hitome Collaboration\ 2018, \pasj, 70, 9
\bibitem[]{}
Ichinohe, Y., Werner, N., Simonescu, A., Allen, S.W., Canning, R.E.A., Ehlert, S., Mernier, F., \& Takahashi, T.\ 2015, \mnras, 448, 2971
\bibitem[]{}
Inoue, H.\ 2014, \pasj, 66, 60
\bibitem[]{}
Jones, C. $\&$ Forman, W. 1984, \apj, 276, 38
\bibitem[]{}
Kaastra, J.S., Ferrigno, C., Tamura, T., Paerels, F.B.S., Peterson, J.R., \& Mittaz, J.P.D. 2001, \aap, 365, L99
\bibitem[]{}
Makishima, K. et al.\ 2001, \pasj, 53, 401
\bibitem[]{}
Markevitch, M., \& Vikhlinin, A.\ 2007, Physics Reports, 443, 1
\bibitem[]{}
Markevitch, M., Vikhlinin, A., \& Mazzotta, P.\ 2001, \apj, 562, L153
\bibitem[]{}
McDonald, M., Veilleux, S., \& Mushotzky, R.\ 2011, \apj, 731, 33
\bibitem[]{}
McDonald, M., Veilleux, S., \& Rupke, D.S.N.\ 2012, \apj, 746, 153
\bibitem[]{}
McDonald, M., Veilleux, S., Rupke, D.S.N., \& Mushotzky, R.\ 2010, \apj, 721, 1262
\bibitem[]{}
McNamara, B.R. $\&$ Nulsen, P.E. 2012, New Journal of Physics, 14, 055023
\bibitem[]{}
Ostriker, E.C.\ 1999, \apj, 513, 252
\bibitem[]{}
Peterson, J.R. et al. 2001, \aap, 365, L104
\bibitem[]{}
Sanders, J.S., Fabian, F.C., \& Taylor, G.B.\ 2005, \mnras, 356, 1022
\bibitem[]{}
Sutherland, R.S., \& Dopita, M.A.\ 1993, \apjs, 88, 253
\bibitem[]{}
Tamura, T. et al. 2001, \aap, 365, L87
\bibitem[]{}
Ueda, S., Kitayama, T., \& Dotani, T.\ 2017, \apj, 837, 34
\bibitem[]{}
ZuHone, J.A., Markevitch, M., \& Johnson, R.E.\ 2010, \apj, 717, 908

\end{thebibliography}
\end{document}